\documentclass[aps,preprint,epsfig,rotate]{revtex4}
\usepackage{graphicx}
\usepackage{bm}
\usepackage{epsfig}

\input epsf
\begin{document}
\title{New method for highly accurate calculations of the photodetachment cross-sections of the negatively charged hydrogen ions}

 \author{Alexei M. Frolov}
 \email[E--mail address: ]{afrolov@uwo.ca}

\affiliation{Department of Applied Mathematics \\
 University of Western Ontario, London, Ontario N6H 5B7, Canada}

\date{\today}

\begin{abstract}

New approach to highly accurate calculations of the photodetachment cross-sections of the negatively charged hydrogen ions is developed. This effective, fast and numerically 
stable method is based on the use of the Rayleigh's formula for the spherical Bessel functions. Photodetachment cross-sections of the negatively charged hydrogen ion(s) 
${}^{\infty}$H$^{-}$, ${}^{1}$H$^{-}$ (protium), ${}^{2}$H$^{-}$ (deuterium D$^{-}$) and ${}^{3}$H$^{-}$ (tritium T$^{-}$) are determined with the use of highly accurate, 
truly correlated, variational wave functions constructed for these ions. Our method allows one to investigate the problem of photodetachment of the negatively charged hydrogen 
ions completely, i.e. without any limitation by numerical values of the photo-electron momentum $p_e$. 

\end{abstract}

\maketitle
\newpage

\section{Introduction}

The negatively charged hydrogen ions ${}^{\infty}$H$^{-}$ (nuclear mass is infinite), ${}^{1}$H$^{-}$ (protium) and ${}^{2}$H$^{-}$ (deuterium, or D $^{-}$) are of great interest in 
stellar astrophysics \cite{Sob} - \cite{Zir}. These ions determine the absorption of infrared and visible radiation in photospheres of all stars, if temperatures in their photospheres 
are restricted between $T_{max} \approx$ 8,250 $K$ (late A-stars) and $T_{min} \approx$ 2,750 $K$ (early M-stars). In the late F, G and early K stars the absorption of infrared and 
visible radiation by the negatively charged hydrogen ions is maximal. This includes our Sun which is a star of spectral type G2. The great role of the H$^{-}$ ions for our Sun was 
suggested by R. Wild in 1939 (see discussions and references in \cite{Sob}, \cite{Aller} and \cite{Zir}). An effective absorbtion of large amount of infrared solar radiation by the 
negatively charged hydrogen ions is crucial to reach a correct thermal balance at our planet. The leading contribution to the light absorption by the negatively charged hydrogen ion 
H$^{-}$ is related to the photodetachment of this ion
\begin{equation}
 {\rm H}^{-} + h \nu = {\rm H}(ns) + e^{-} \label{e0}
\end{equation}
where the notation $h \nu$ designates the incident light quantum, while $e^{-}$ means the free electron, or photo-electron, for short. The notation H($ns$) means the hydrogen atom in 
the bound $ns-$state, where $n$ is the principal quantum number and $s$ means $s-$states with $\ell = 0$ where $\ell$ is the angular momentum of the hydrogen atom. The main goal of this 
study is to determine and investigate the photodetachment cross-sections of the negatively charged hydrogen ion(s). A large amount of preliminary work for this problem, including 
derivation of different formulas needed for numerical calculations of the photodetachment cross-sections of the H$^{-}$ ions can be found in \cite{Fro2014}. The paper \cite{Fro2014} 
also contains a review of the photodetachment of negatively charged hydrogen ions and extensive bibliography on this subject. This allows us to restrict our analysis in this study to 
the discussion of computational method and results obtained with the use of this method. Note also that a very good discussion of the photodetachment of the negatively charged hydrogen 
ions can be found in some textbooks, e.g., in \cite{FrFish}, \cite{AB} and \cite{BS}. 

Our approach developed in this study is based on the use of highly accurate, three-body (or two-electron) wave functions constructed for the negatively charged hydrogen ions. Such wave 
functions are truly correlated. The final state wave functions are represented in the form of products of two single-electron wave functions, i.e. they are not correlated wave functions. 
In general, the overlap integrals between truly correlated and non-correlated wave functions cannot be computed without a substantial loss of accuracy. For the photodetachment of 
the negatively charged hydrogen ions this problem, i.e. loss of overall accuracy, is known since the first works by Chandrasekhar \cite{Chand} - \cite{Chand2}. In this study we avoid 
any loss of high accuracy during calculations of the three-particle integrals with spherical Bessel/Neumann functions by using our recently developed, original method for computation of 
such integrals written in relative coordinates. This method allows one to perform accurate numerical calculations of the photodetachment cross-sections for different hydrogen isotopes 
and different final states in the hydrogen atom(s). In such calculations we found no limiting values for the momenta $p_e$ of photo-electron. In other words, all three-particle integrlas 
with spherical Bessel functions absolutely converge in this new method. Note also that in contrast with other similar methods in our approach we operate only with regular three-particle 
integrlas with Bessel functions, i.e. there is no need to discuss convergence of the Frullanian and/or truly singular integrals.     

This paper has the following structure. First, in the next Section we discuss the general formulas for the differential and total photodetachment cross-sections of the negatively charged 
hydrogen ion. It is shown that calculation of the photodetachment cross-section is reduced to numerical calculations of three-particle integrals with the spherical Bessel functions. Our 
approach applied in this study for calculations of such integrals is based on the Rayleigh's formula for spherical Bessel function is discussed in Section III. Analytical formulas for the 
three-particle integrals with spherical Bessel functions are derived in Section IV. In Section V we discuss numerical computations of the photodetachment cross-sections of the negatively 
charged hydrogen ion(s). Concluding remarks can be found in the last Section.

\section{Photodetachment cross-section}

In \cite{Fro2014} we have derived the following formula for the differential cross-section of the photodetachment of the negatively charged hydrogen ion 
\begin{eqnarray}
 d\sigma_{ph} = 8 \pi \alpha a^2_0 {\cal K} \Bigl(\frac{p_e}{\frac{p^2_e}{2} + I_{n}}\Bigr) ({\bf n}_e \times {\bf k}_f)^2 \Bigl| {\cal R}_{i \rightarrow f} \Bigr|^2 do \label{difcros} 
\end{eqnarray}
where ${\cal K}$ is some numerical constant defined below, $\alpha = \frac{e^2}{\hbar c} = 7.2973525698 \cdot 10^{-3} \Bigl(\approx \frac{1}{137}\Bigr)$ is the dimensionless fine structure 
constant and $a_0 \approx 0.52917721092 \cdot 10^{-8}$ $cm$ is the Bohr radius (unless otherwise specified all numerical values of physical constants in this study are taken from \cite{CRC}). 
The ${\bf k}_f$ vector is the unit vector which describes the photon propagation. Analogously, the unit vector ${\bf n}_e = \frac{{\bf p}_e}{p_e}$ describe propagation of the free photo-electron 
which arises during photodetachment. Also, in this formula $p_e = \mid {\bf p}_e \mid$ is the momentum (absolute value) of the photo-electron and $I_{n}$ is the ionization potential of the 
negatively charged hydrogen ion determined in the case when the final hydrogen atom arises in the bound $ns-$state, where $n \ge 1$ and symbol $s$ means that the angular momentum of this state 
equals zero, i.e. $\ell = 0$. The notation ${\cal R}_{i \rightarrow f}$ in this formula is the radial part of the photodetachment amplitude. From this formula one finds the following formula for 
the photodetachment cross-section $\sigma_{ph}$ of the negatively charged hydrogen ion(s):
\begin{eqnarray}
 \sigma_{ph} &=& \frac{64 \pi^2}{3} \alpha a^2_0 {\cal K} \Bigl(\frac{p_e}{\frac{p^2_e}{2} + I_{n}}\Bigr) \Bigl| {\cal R}_{i \rightarrow f} \Bigr|^2 = \frac{64 \pi^2}{3} \cdot 2.043466844
 \cdot 10^{-19} \cdot {\cal K} \Bigl(\frac{p_e}{\frac{p^2_e}{2} + I_{n}}\Bigr) \Bigl| {\cal R}_{i \rightarrow f} \Bigr|^2 \; \; \; cm^{2} \nonumber \\
 &=& 33.9954673 \cdot 10^{-18} \cdot {\cal K} \Bigl(\frac{p_e}{\frac{p^2_e}{2} + I_{n}}\Bigr) \Bigl| {\cal R}_{i \rightarrow f} \Bigr|^2 \; \; \; cm^{2} \; \; \; \label{sigtot} 
\end{eqnarray}
For truly correlated three-particle wave functions the explicit formula for the radial part of the photodetachment amplitude is 
\begin{eqnarray}
 & & {\cal R}_{i \rightarrow f} = \frac12 \Bigl[ \sum^{N}_{j=1} C_j \Bigl[\alpha_j \int_{0}^{+\infty} \int_{0}^{+\infty} \int^{r_{31} + r_{32}}_{\mid r_{31} - r_{32} \mid} R_{n0}(r_{31}) 
 j_{1}(p_e r_{32}) \times \nonumber \\
 & & \exp(-\alpha_j r_{32} -\beta_j r_{31} - \gamma_j r_{21}) r_{32} r_{31} r_{21} dr_{32} dr_{31} dr_{21} + \beta_j \times \label{int0} \\
 & & \int_{0}^{+\infty} \int_{0}^{+\infty} \int^{r_{31} + r_{32}}_{\mid r_{31} - r_{32} \mid} R_{n0}(r_{31}) j_{1}(p_e r_{32}) \exp(-\beta_j r_{32} -\alpha_j r_{31} - \gamma_j r_{21}) r_{32} 
 r_{31} r_{21} dr_{32} dr_{31} dr_{21} \Bigr] \nonumber
\end{eqnarray}
This amplitude is the overlap of the incident wave function of the H$^{-}$ ion (truly correlated) and the final state wave function which is the product of the radial part of the total wave function 
$R_{n0}(r_{31})$ of the final hydrogen atom H in one of its bound $ns-$states and the wave function of the moving photo-electron. For the ground state in the final hydrogen atom one finds 
$R_{n0}(r_{31}) = 2 \exp(- r_{31})$. If we can neglect by any interaction between this photo-electron and neutral hydrogen atom, then the wave function of the free photo-electron is a plane wave, 
which is represented as a combination of products of the spherical Bessel functions of the first kind with the corresponding Legendre polynomials (see, e.g., \cite{Rose}, p.91). This formula is 
called the Rayleigh expansion of the plane wave \cite{Rose}. Conservation of the angular momentum during photodetachment leads to the selection of only one $\ell-$component in such an expansion.    

It is shown below that with our choice of the incident and final wave functions the factor ${\cal K}$ in Eqs.(\ref{difcros}) - (\ref{sigtot}) equals $\frac{1}{8 \pi^2}$. In this case the formulas  
Eqs.(\ref{difcros}) - (\ref{sigtot}) are written in the form
\begin{eqnarray}
 d\sigma_{ph} = \frac{1}{\pi} \alpha a^2_0 \Bigl(\frac{p_e}{\frac{p^2_e}{2} + I_{n}}\Bigr) ({\bf n}_e \times {\bf k}_f)^2 \Bigl| {\cal R}_{i \rightarrow f} \Bigr|^2 do &=&
 6.504556985 \cdot 10^{-20} \Bigl(\frac{p_e}{\frac{p^2_e}{2} + I_{n}}\Bigr) ({\bf n}_e \times {\bf k}_f)^2 \times \nonumber \\ 
 & & \Bigl| {\cal R}_{i \rightarrow f} \Bigr|^2 do \label{difcros1} 
\end{eqnarray}
and 
\begin{eqnarray}
 \sigma_{ph} = \frac{8}{3} \alpha a^2_0 \Bigl(\frac{p_e}{\frac{p^2_e}{2} + I_{n}}\Bigr) \Bigl| {\cal R}_{i \rightarrow f} \Bigr|^2 = 5.44924918
 \cdot 10^{-19} \Bigl(\frac{p_e}{\frac{p^2_e}{2} + I_{n}}\Bigr) \Bigl| {\cal R}_{i \rightarrow f} \Bigr|^2 \; \; \; cm^{2} \label{sigtt}
\end{eqnarray}
These formulas are used in all calculations performed for this study. In the next two Sections we derive analytical formulas for the matrix element ${\cal R}_{i \rightarrow f}$ which is included 
in the last two formulas.

\section{Rayleigh's formula for spherical Bessel functions}

The method described below is, probably, the most effective and accurate approach which is used to determine the photodetachment cross-section of the negatively charged hydrogen ion(s). High 
efficiency and accuracy of this method directly follows from its simplicity and transparency. The method is based on the Rayleigh's formula for the Bessel functions \cite{AS}. In the case 
of the $j_{1}(z) = j_{1}(p r_{32})$ Bessel function the Rayleigh's formula takes the form 
\begin{equation}
 j_{1}(z) = - \frac{d j_{0}(z)}{d z} = - \frac{d}{d z} \Bigl( \frac{\sin z}{z} \Bigr) \; \; \; . \; \; \; \label{eq1}
\end{equation}
For $z = p r_{32}$ one finds from this equation 
\begin{equation}
 j_{1}(p r_{32}) = -\frac{1}{p^{2}} \frac{\partial}{\partial r_{32}} \Bigl( j_{0}(p r_{32}) \Bigr) = -\frac{1}{p^{2}} \Bigl[\frac{\partial}{\partial r_{32}} 
 \Bigl( \frac{\sin (p r_{32})}{r_{32}} \Bigr) \Bigr] \; \; \; . \; \; \; \label{eq2}
\end{equation}
where $p = p_e$ is the absolute value of momentum of the outgoing photo-electron ${\bf p}_e$, while $r_{32}$ is the scalar distance between the hydrogen nucleus (heavy particle 3)
and photo-electron (particle 2). The particle 1 designates another (first) electron which remains bound after photodetachment. The same system of notations is used everywhere below in this 
study.

Let us designate the wave function of the incident H$^{-}$ ion in the ground $1^1S-$state by $\Psi_{{\rm H}^{-}}(r_{32}, r_{31}, r_{21})$, while the notation $\phi_n(r_{31})$ stands for the wave 
function of the final $ns-$state of the hydrogen atom. Then we can write the following expression for the amplitude of the photodetachment $M_{i \rightarrow f}$ \cite{Fro2014}:
\begin{eqnarray}
 M_{i \rightarrow f} &=& \int \phi_n(r_{31}) j_{1}(p r_{32}) \Psi(r_{32}, r_{31}, r_{21}) dV \nonumber \\ 
 &=& - \frac{1}{p^2} \int \phi_n(r_{31}) \Bigl[\frac{\partial}{\partial r_{32}} j_{0}(p r_{32}) \Bigr] \Psi_{{\rm H}^{-}}(r_{32}, r_{31}, r_{21}) dV \; \; \; . \; \; \; \label{eq3}
\end{eqnarray}  
where $dV = r_{32} r_{31} r_{21} dr_{32} dr_{31} dr_{21}$ is an elementary volume in the relative coordinates $r_{32}, r_{31}$ and $r_{21}$. Each of the relative coordinates is the difference 
between the two Cartesian coordinates of the corresponding particles, e.g., $r_{ij} = \mid {\bf r}_i - {\bf r}_j \mid$. It follows from such a definition that: (a) relative coordinates are 
symmetric, i.e. $r_{ij} = r_{ji}$, and (b) the following inequalities are always obeyed: $\mid r_{ik} - r_{jk} \mid \le r_{ij} \le r_{ik} + r_{jk}$ for $( i, j, k) = ( 1, 2, 3)$. Now, we can 
transform the expression in the right-hand side of Eq.(\ref{eq3}) to the following form
\begin{equation}
 M_{i \rightarrow f} = \frac{1}{p^2} \int \phi_n(r_{31}) j_{0}(p r_{32}) \Bigl[\frac{\partial}{\partial r_{32}} \Phi(r_{32}, r_{31}, r_{21}) \Bigr] dV 
 \; \; \; . \; \; \; \label{eq4}
\end{equation} 
which contains the partial derivative of the incident wave function, i.e. the bound state wave function of the H$^{-}$ ion. The function $\Phi(r_{32}, r_{31}, r_{21})$ in Eq.(\ref{eq4}) is related 
with the actual wave functions of the negatively charged hydrogen ion $\Psi_{{\rm H}^{-}}(r_{32}, r_{31}, r_{21})$ by the following relation
\begin{eqnarray}
  \Phi(r_{32}, r_{31}, r_{21}) = \frac{\partial}{\partial r_{32}} \Psi_{{\rm H}^{-}}(r_{32}, r_{31}, r_{21})
\end{eqnarray}

In this study the trial variational wave functions of the ground (singlet) $1^1S-$state of the negatively charged hydrogen ions are constructed in the following (exponential) form (see also 
\cite{Fro2014}) 
\begin{eqnarray}
  \Psi_{{\rm H}^{-}} &=& \frac12 ( 1 + \hat{P}_{12} ) \sum^{N}_{i=1} C_i \exp(-\alpha_i r_{32} - \beta_i r_{31} - \gamma_i r_{21}) \nonumber \\
  &=& \frac12 ( 1 + \hat{P}_{12} ) \sum^{N}_{i=1} C_i \exp[-(\alpha_{i} + \beta_{i}) u_{3} - (\alpha_{i} + \gamma_{i}) u_{2} - (\beta_{i} + \gamma_{i}) u_{3})] \; \; \; \label{exp}
\end{eqnarray}
where the notation $\hat{P}_{12}$ stands for the permutation operator of identical particles (electrons), $C_i$ ($i = 1, 2, \ldots, N$) are the linear parameters of the exponential expansion, 
Eq.(\ref{exp}), while $\alpha_i, \beta_i$ and $\gamma_i$ are the non-linear parameters of this expansion. The trial wave function, Eq.(\ref{exp}), has the correct permutation symmetry which is 
required for the spatial part of the singlet wave functions in the two-electron atomic systems. The non-linear parameters in Eq.(\ref{exp}) must be varied in calculations to increase the overall 
efficiency and accuracy of the method. There are $3 N$ additional conditions for these non-linear parameters are: $\alpha_{i} + \beta_{i} > 0, \alpha_{i} + \gamma_{i} > 0, \beta_{i} + \gamma_{i} 
> 0$ ($i = 1, 2, \ldots, N$). These conditions must always be obeyed to guarantee convergence of all three-particle integrals arising in computations. In this study we assume that all non-linear 
parameters in Eq.(\ref{exp}) are the real numbers with zero complex parts.  

Note that Eq.(\ref{exp}) is, in fact, the Laplace transform of the wave function $\Psi$. It follows form Eq.(\ref{exp}) that  
\begin{eqnarray}
 \frac{\partial}{\partial r_{32}} \Psi_{{\rm H}^{-}}(r_{32}, r_{31}, r_{21}) &=& \frac12 \sum^{N}_{i=1} C_i \Bigl[ \alpha_i \exp(-\alpha_i r_{32} - \beta_i r_{31} - \gamma_i r_{21}) \nonumber \\
  &+& \beta_i \exp(-\beta_i r_{32} - \alpha_i r_{31} - \gamma_i r_{21}) \Bigr] \label{deriv1}
\end{eqnarray}
and 
\begin{eqnarray}
 \frac{\partial^{2}}{\partial r^{2}_{32}} \Psi_{{\rm H}^{-}}(r_{32}, r_{31}, r_{21}) &=& \frac12 \sum^{N}_{i=1} C_i \Bigl[ \alpha^{2}_i \exp(-\alpha_i r_{32} - \beta_i r_{31} - \gamma_i r_{21}) 
 \nonumber \\
 &+& \beta^{2}_i \exp(-\beta_i r_{32} - \alpha_i r_{31} - \gamma_i r_{21}) \Bigr] \label{deriv2}
\end{eqnarray}
The formulas, Eqs.(\ref{deriv1}) - (\ref{deriv2}), are used in this study to determine the first and second partial derivatives of the trial wave functions. 

\section{Analytical formulas for three-particle integrals with Bessel functions}

In this Section we discuss analytical approach which was found to be very effective for analytical and numerical computations of the three-particle integral(s) derived above (see, Eq.(\ref{int0}))
which include one spherical Bessel function $j_{0}(p r_{32})$. The three-particle integral with the spherical Bessel function $j_{0}(p r_{32})$ which is needed in such calculations takes the form 
\begin{eqnarray}
 I(\alpha, \beta, \gamma; p) &=& \frac{1}{p^2} \int \int \int \exp(-\alpha_i r_{32} - \beta_i r_{31} - \gamma_i r_{21}) j_{0}(p r_{32}) r_{32} r_{31} r_{21} dr_{32} dr_{31} dr_{21} \nonumber \\
 &=& \frac{1}{p^2} \int \int \int \exp(-\alpha_i r_{32} - \beta_i r_{31} - \gamma_i r_{21}) \sin(p r_{32}) r_{31} r_{21} dr_{32} dr_{31} dr_{21} \label{j0}
\end{eqnarray} 
It is clear that analytical and/or numerical computations of three-particle integrals written in the relative coordinates is a difficult problem, since the three relative coordinates $r_{32}, 
r_{31}$ and $r_{21}$ are not truly independent. Indeed, three inequalities $\mid r_{ik} - r_{jk} \mid \le r_{ij} \le r_{ik} + r_{jk}$ are always obeyed for three relative coordinates $r_{32}, r_{31}$ 
and $r_{21}$ (they are mentioned in the previous Section). In real three-body calculations it is better to use three perimetric coordinates $u_1, u_2, u_3$ which are simply related with the perimetric 
coordinates: $u_i = \frac12 (r_{ij} + r_{ik} - r_{jk})$, where $(i, j, k) = (1, 2, 3)$ and $r_{ij} = r_{ji}$. The inverse relations are: $r_{ij} = u_{i} + u_{j}$. Each of the three perimetric 
coordinates $u_i (i = 1, 2, 3)$ is: (1) non-negative, (2) truly independent from other perimetric coordinates, and (3) changes from zero to infinity. In perimetric coordinates the last integral from 
Eq.(\ref{j0}) is written in the form
\begin{eqnarray}
 I(\alpha_i, \beta_i, \gamma_i; p) &=& \frac{2}{p^2} \int^{+\infty}_{0} \int^{+\infty}_{0} \int^{+\infty}_{0} \exp[-(\alpha_i + \beta_{i}) u_3 - (\alpha_i + \gamma_i) u_{2} - 
 (\beta_i + \gamma_i) u_{1}] \times \nonumber \\
 & & \sin(p u_{2} + p u_{3}) (u^{2}_{1} + u_{1} u_{2} +  u_{1} u_{3} +  u_{2} u_{3}) du_{1} du_{2} du_3 \label{j01}
\end{eqnarray} 
where the factor 2 is the Jacobian of the transition from the relative to perimetric coordinates $(r_{32}, r_{31}, r_{21}) \rightarrow (u_1, u_2, u_3)$. To derive the final expression we apply the known 
trigonometric formula $\sin(p u_{2} + p u_{3}) = \sin(p u_{2}) \cos(p u_{3}) + \cos(p u_{2}) \sin(p u_{3})$ and introduce the following notations: $Z = \alpha_i + \beta_{i}, Y = \alpha_i + \gamma_i, X = 
\beta_i + \gamma_i$. In these notations the integral Eq.(\ref{j01}) is written as the sum of eight integrals. 

To illustrate the process of integration in perimetric coordinates let us consider analytical computations of one of these eight integrals, e.g., 
\begin{eqnarray}
 I_3 &=& \frac{2}{p^2} \int^{+\infty}_{0} \int^{+\infty}_{0} \int^{+\infty}_{0} \exp[-Z u_3 -Y u_{2} - X u_{1}] u_{1} u_{2} \sin(p u_{2}) \sin(p u_{3}) du_{1} du_{2} du_3 \nonumber \\
 &=& \frac{2}{p^2} \int^{+\infty}_{0} \exp[- X u_{1}] u_{1} du_{1} \int^{+\infty}_{0} \exp[-Y u_{2}] u_{2} \sin(p u_{2}) du_{2} \int^{+\infty}_{0} \exp[-Z u_3] \sin(p u_{3}) du_3 \nonumber \\
 &=& \frac{2}{p^2} \cdot \frac{1}{X^2} \cdot \frac{2 p Y}{[Y^{2} + p^{2}]^2} \cdot \frac{Z}{Z^{2} + p^{2}} = \frac{2 Y Z}{p X^2 [Y^{2} + p^{2}]^{2} (Z^{2} + p^{2})} \label{I3}
\end{eqnarray}
Calculation of seven remaining integrals is also simple and transparent. Note that our analytical computations are based on the formulas from \cite{GR}. The final expression for the integral $I(\alpha_i, 
\beta_i, \gamma_i; p)$ takes the form
\begin{eqnarray}
 I(\alpha_i, \beta_i, \gamma_i; p) &=& \frac{2}{p X (Y^{2} + p^{2}) (Z^{2} + p^{2})} \Bigl[ \frac{2 (Z + Y)}{X^{2}} + \frac{2 Y Z}{X (Y^{2} + p^{2})} + \frac{2 Y Z}{X (Z^{2} + p^{2})} \label{j11} \\ 
 &+& \frac{Y^{2} - p^2}{X (Y^{2} + p^{2})} + \frac{Z^{2} - p^2}{X (Z^{2} + p^{2})} + \frac{2 Y (Z^{2} - p^2)}{(Y^{2} + p^{2}) (Z^{2} + p^{2})} + \frac{2 Z (Y^{2} - p^2)}{(Y^{2} + p^{2}) (Z^{2} + p^{2})}
 \Bigr] \nonumber
\end{eqnarray} 
Note that the formula, Eq.(\ref{j11}), does not contain any singular and/or quasi-singular term represented by the Frullanian integrals \cite{Frul}, \cite{Jef} which must be regularized before actual 
numerical computations. This is an obvious advantage of our approach which is based on the Rayleigh's formula for spherical Bessel functions. It should be mentioned here that integrals of the general type 
$\int_{0}^{\infty} \frac{f(a x) - f(b x)}{x} dx$ were described by Italian mathematician G. Frullani in 1828 \cite{Frul}. These integrals look like singular, but they converge, if the finite values $f(0) 
(\ne 0)$ and $f(\infty)$ do exist and $a$ and $b$ are both positive (or negative) real numbers. In the photodetachment of the H$^{-}$ ion one always finds three-particle Frullanian integrals which look like
singular integrals, but can be reduced to a few infinite (but convergent!) sums of the regular three-particle integrals.     

In earlier studies \cite{FrWa2014}, \cite{FrWa2015} we have developed a number of different methods (six methods) for analytical and/or numerical computations of three-particle integrals which 
contain Bessel functions of the first and second kind. Some of these methods are more successful in applications than others. In particular, we have found that methods based on power-series expansions 
of the spherical Bessel functions are not appropriate for highly accurate computations of the overlap integrals with the bound state wave functions. The reason is obvious, since contribution of the 
three-particle integrals rapidly increases when powers of the relative coordinate $r_{32}$ increase in such integrals. In general, any power-series expansion of the spherical Bessel functions contains 
large and very large powers of this relative coordinate $r_{32}$. Convergence rate of such three-particle integrals slows down rapidly. For large $n$ all integrals which contain $r^{n}_{32}$ become 
divergent, if the contribution of high power of $r_{32}$ cannot be compensated by the increasing powers of some small parameter(s). This explains a restricted use of the power-series expansion for the 
spherical Bessel functions in actual computations. Our method developed in this study is free form these difficulties. Moreover, the formulas presented above are remarcably stable in calculations with large 
number of the non-linear parameters. This allows us to solve the problem to very high accuracy. Furthermore, by using these formulas we can also determine the photodetachment cross-section in those cases 
when the final hydrogen atom is formed in the excited states. The power-type series for Bessel functions derived in \cite{FrWa2014} and \cite{FrWa2015} are used in calculations of the three-particle 
integrals with Bessel functions only to check our current formulas for small values of $p_e (\le 0.015 - 0.030)$.    

\section{Numerical calculations}

Numerical calculations of the photodetachment cross-section of the negatively charged hydrogen ions H$^{-}$ are performed in this study with the use of different variational wave functions, or trial wave 
functions, for short. These trial wave functions contain different number of the basis wave functions, Eq.(\ref{exp}), i.e. they have different overall accuracy. In this study we have used trial wave 
functions, Eq.(\ref{exp}), with $N$ = 50, 100, 200, 300 and 350 basis functions (exponents). Convergence of our photodetachment cross-section upon the number of basis functions $N$ in the trial wave 
functions is shown in Table I for the H$^{-}$ ion with infinitely heavy nuclear mass which is designated as the ${}^{\infty}$H$^{-}$ ion. 

Table II contains photodetachment cross-sections of the ${}^{\infty}$H$^{-}$ ion computed for 200 different values of $p = p_e$, which is the absolute value of momentum ${\bf p}_e$ of the emitted 
photo-electron. In our Tables the kinetic energy of the photo-electron $E_e = \frac{p_e^2}{2}$ is used, since the dependence $\sigma_p(E_e)$ is directly measured in modern experiments. In Table II the 
kinetic energy of the photo-electron $E_e$ is related with the `index' $i$ as follows: 
\begin{eqnarray}
    E_e(i) = \frac{p^2_i}{2} \; \; \; , \; \; \; where \; \; \; p_i = 0.005 + (i - 1) \cdot 0.005
\end{eqnarray}
in atomic units. The photodetachment cross-section of the H$^{-}$ ion is written in the form:
\begin{eqnarray}
 \sigma_{ph} = \frac{64 \pi^2}{3} \cdot \alpha a^{2}_0 \cdot {\cal K} \cdot \Bigl(\frac{\sqrt{2 E_e}}{E_e + I_1}\Bigr) \Bigl| {\cal R}_{i \rightarrow f} \Bigr|^2 \; \; \; cm^{2} \; \; \; \label{sigmt} 
\end{eqnarray}
where ${\cal R}_{i \rightarrow f}$ is the radial part of the photodetachment amplitude, Eq.(\ref{int0}) and $I_1 = E({}^{\infty}$H$^{-})$ + 0.5 $a.u.$, where $E({}^{\infty}$H$^{-})$ is the total energy of the 
H$^{-}$ ion. The numerical value of the factor ${\cal K}$ in this formula equals $\frac{1}{8 \pi^2}$ which equals to the square of the norm of the angular part of the truly correlated, two-electron wave 
function of the ${}^{infty}$H$^{-}$ ion. Finally, one finds the following expression for the photodetachment cross-section     
\begin{eqnarray}
 \sigma_{ph} &=& \frac{8}{3} \alpha a^{2}_{0} \cdot \Bigl(\frac{\sqrt{2 E_e}}{E_e + I_1}\Bigr) \Bigl| {\cal R}_{i \rightarrow f} \Bigr|^2 \; \; \; cm^{2} \; \; \; \label{sigmtt} \\
  &=& 5.449249176 \cdot 10^{-19} \cdot \Bigl(\frac{\sqrt{2 E_e}}{E_e + I_1}\Bigr) \Bigl| {\cal R}_{i \rightarrow f} \Bigr|^2 \; \; \; cm^{2} \nonumber 
\end{eqnarray}
where the expression for ${\cal R}_{i \rightarrow f}$ is given by Eq.(\ref{int0}). The formula, Eq.(\ref{sigmtt}), has been used in our calculations of the photodetachment cross-section of the H$^{-}$ ion
(see Tables I and II). In general, our results from Tables I and II agree very well with the results of the earlier and recent experimental studies \cite{PR1955A} - \cite{PRA2015}. Agreement between our 
photodetachment cross-section of the H$^{-}$ ion and results obtained in other theoretical/computational works (see, e.g., \cite{PRA1994}) can be considered as very good for those cases when the final hydrogen 
atom is formed in its ground $1s-$state. Note that experimental studies of the H$^{-}$ ion photodetachment with the formation of the final hydrogen atom in the excited $S-$states are extremely difficult to 
perform. Agreement of our photodetachment cross-sections determined in the case of $2s$ hydrogenic state and results known from earlier theoretical studies \cite{BD1} and \cite{excite} is good. However, it is 
clear that to obtain a perfect agreement between theoretical cross-sections in this case one needs to conduct additional research and evaluations (see also discussion in the Conclusion).   
   
Another interesting question is to determine photodetachment cross-section of the H$^{-}$ ion in the case when the final hydrogen atom arises in the first excited $s-$state. All calculations are almost 
identical to the case considered above when the final hydrogen atom is formed in its ground $1^2S-$state. However, in this case the ionization potential of the H$^{-}$ ion is $I = E($H$^{-}) + 0.25$ 
$a.u.$ and in the formula for the photodetachment amplitude one needs to replace the bound state wave functions of the hydrogen atom, e.g., for the ${}^{\infty}$H atom:
\begin{eqnarray}
  2 \cdot \exp(- r) \rightarrow \frac{1}{\sqrt{2}} \exp(- \frac{r}{2}) \Bigl(1 - \frac{r}{2}\Bigr)  \; \; \; \label{wavefun} 
\end{eqnarray}
The photodetachment cross-section of the ${}^{\infty}$H$^{-}$ ion in the case when the final hydrogen atom arises in the first excited $s-$state is determined with the same trial wave functions of
the H$^{-}$ ion (see above, for $N$ = 350). The photodetachment cross-section (in $cm^{2}$) for this case is presented in Table III. 

Note that calculations of the photodetachment amplitude in the case of the first excited state (in the final hydrogen atom) requires computation of the two three-particle integrals with Bessel functions. One 
of these integrals essentially coincides with Eq.(\ref{j0}), while another integral contains an extra power of $r_{31}$, i.e. 
\begin{eqnarray}
 J(\alpha, \beta, \gamma; p) = \frac{1}{p^2} \int \int \int \exp(-\alpha_i r_{32} - \beta_i r_{31} - \gamma_i r_{21}) \sin(p r_{32}) r^{2}_{31} r_{21} dr_{32} dr_{31} dr_{21} \label{j2}
\end{eqnarray} 
Analytical expression for this integral can be derived in the same way as we did to determine the integral $I(\alpha, \beta, \gamma; p)$, Eq.(\ref{j0}). However, the final expression for the $J(\alpha, 
\beta, \gamma; p)$ integral is difficult, since it contains a large number of different terms. To determine the integral $J(\alpha, \beta, \gamma; p)$ in this study we apply another method which is based on 
the use of numerical differentiation of the integral $I(\alpha, \beta, \gamma; p)$, Eq.(\ref{j0}), upon the non-linear parameter $\alpha$. To calculate this first order derivative we use the following formula
\begin{eqnarray}
 J(\alpha, \beta, \gamma; p) &=& -\frac{\partial}{\partial \alpha} I(\alpha_i, \beta_i, \gamma_i; p) \approx \frac{1}{12 \Delta} \Bigl[ I(\alpha_i - 2 \Delta, \beta_i, \gamma_i; p) - 8 I(\alpha_i - \Delta, 
 \beta_i, \gamma_i; p) \nonumber \\
 &+& 8 I(\alpha_i + \Delta, \beta_i, \gamma_i; p) - I(\alpha_i + 2 \Delta, \beta_i, \gamma_i; p) \Bigr] \label{derv1}
\end{eqnarray}
where the three-particle integral $I(\alpha_i, \beta_i, \gamma_i; p)$ is defined by Eq.(\ref{j01}). In Eq.(\ref{derv1}) for each $\alpha_i$ the parameter $\Delta$ is defined as follows: $\Delta = \delta \cdot 
\alpha_i$, where $i = 1, \ldots, N$ and $\delta$ is a small number, e.g., $\delta = 1 \cdot 10^{-6}$, or $\delta = 1 \cdot 10^{-7}$. The overall accuracy of the formula, Eq.(\ref{derv1}), can be evaluated as 
$f \cdot \frac{\Delta^4}{90}$ (see, e.g., \cite{Num}), where $f$ is a positive numerical factor. To prove the correctness of numerical calculations of the first-order derivative one needs to perform a series 
of calculations by varying small parameter $\delta$ from, e.g., $1 \cdot 10^{-4}$ up to $1 \cdot 10^{-8}$ (if calculations are performed with with the use of the quadruple precision). Such a series of 
calculations allows one to evaluate the total number of stable decimal digits in this derivative. In general, the computed derivative must be stable in 10 - 12 decimal digits (at least).

As follows from comparison of Tables II and III the photodetachment cross-section for the ground state in the final hydrogen atom is significantly larger than analogous cross-section for the first excited 
state. This can be explained by `almost exact' orthogonality of the two-electron wave function of the H$^{-}$ ion $\Psi_{{\rm H}^{-}}(r_{32}, r_{31}, r_{21})$ and the $2s-$hydrogenic orbital $\phi(r_{31})$. 
This fact is known for the natural orbital expansions constructed for the negatively charged atomic ions, e.g., for the hydrogen ion(s) H$^{-}$. It is closely related to the fact that the negatively charged 
hydrogen ion has only one bound $1^1S-$state. However, there is no such an `almost exact' orthogonality between the $\Psi_{{\rm H}^{-}}(r_{32}, r_{31}, r_{21})$ function and $3s-$ and $4s-$hydrogenic orbitals. 
This means that the corresponding photodetachment cross-sections for the $3s-$ and $4s-$states in the final hydrogen atom can be larger than for the $2s-$state. Another interesting detail which has been noticed 
in our calculations is the presence of multiple maxima in photodetachment cross-sections of all excited $ns-$states for $n \ge 3$ (in the final hydrogen atom). Already, for the $3s-$state in the final hydrogen 
atom one finds two such maxima in the photodetachment cross-section. This follows from the explicit form of the photodetachment amplitude ${\cal R}_{i \rightarrow f}$, Eq.(\ref{int0}) and the fact that the 
hydrogenic $R_{n0}(r)$ functions (as well as spherical Bessel function $j_{1}(p_e r)$) oscillate, i.e. vary between their maximal and minimal values for $r \ge 0$ (and $n \ge 1$). This interesting problem will 
be discussed in our next study. 

The energy conservation law leads to the conclusion that formation of the final hydrogen atom in the excited atomic states, including excited $s-$states, requires substantially larger energy of the incident 
photon. However, if we assume that the actual stellar photosphere is at thermal equilibrium with $T_{max} \approx$ 4000 - 8000 $K$, then the part of such `high-energy photons' with energies $\ge$ 80,000 $K$ is 
extremely small. For instance, to produce photodetachment of the H$^{-}$ ion with the formation of the final hydrogen atom in the excited $2s-$state the `photon temperature' $T_f$ must be $\ge$ 78,944 $K$. 
Analogous `photon temperature' to form the final H atom in the excited $3s-$state is around 118,416 $K$. For our Sun the fraction of such `hot' thermal photons with $T \ge$ 78944 $K$ is very small ($\le 1 
\cdot 10^{-2}$ \%). Even in photospheres of late $A-$stars the fraction of `hot' thermal photons is less than 0.5 \%. In reality, such photons become noticeable in photospheres of B-stars with surface temperature 
$T \ge 20,000 K$. However, in classical B-stars the photodetachment of the H$^{-}$ ions does not play any role and most of the emitted radiation is absorbed by neutral hydrogen helium atoms and helium 
one-electron ions. This situation can be different in the Be-stars, i.e. in the B-stars which also include a rapidly rotating, low-dense disk of hydrogen atoms \cite{Sob}, \cite{Malb}. Spatial radius of such a 
`hydrogen' disk $R_D$ is substantially larger than the radius of the central hot star $R_S$. For a typical Be-star $R_D \approx 150 \cdot 10^{8}$ $km$, while $R_S \le 3 \cdot 10^{6}$ $km$. Illumination of this 
disk by radiation emitted by the central hot B-star produces excitations of hydrogen atoms which are moving in the stellar disk. Excitation of hydrogen atoms and following emission of absorbed radiation generates 
a very special optical spectra of the Be-stars, which contain thermal (continous) spectra of radiation emitted from the central hot B-star and a linear spectra which corresponds to the transition between excited 
states in the hydrogen atoms moving in the stellar disk. A well known Be-star is the $\gamma$ Cassiopeia which can easily be located and observed at the clear night sky in Northern Hemisphere. The emission 
spectrum of the $\gamma$ Cassiopeia contains a typical optical spectrum of a hot B-star and the red line which corresponds to the Balmer optimcal series in hydrogen atoms. Such a very intense red line essentially 
determines the spectrum of this star. In general, $\approx 35$ \% of all B-stars are the Be-stars, i.e. they have an additional low-dense disk which contains hydrogen atoms. For such stars photodetachment of the 
H$^{-}$ ions with the formation of the final hydrogen atoms in their excited states may be important in some outer areas of stellar disk. In other stars the photodetachment of the H$^{-}$ ions which leads to the 
formation of the final hydrogen atoms in the excited states is of only a restricted academic interest.     

Finally, let us consider the photodetachment of the actual hydrogen ions with finite nuclear masses, i.e. ions of the protium ${}^{1}$H$^{-}$, deuterium ${}^{2}$H$^{-}$ and tritium ${}^{3}$H$^{-}$. To consider
photodetachment of these negatively charged ions one needs to take into account the effect of `nuclear recoil', i.e. recoil of the central hydrogen atom. In this study we apply the same nuclear masses of the 
hydrogen isotopes which were used in \cite{Fro2014}:
\begin{eqnarray}
  M_p = \frac{938.272046}{0.510998910} \; \; , \; \; M_d = \frac{1875.612859}{0.510998910} \; \; , \; \; M_t = \frac{2808.920906}{0.510998910} \; \; , \; \; \label{masses} 
\end{eqnarray}
These nuclear masses are the most recent values obtained in high-energy experiments. The best-to-date total energies obtained for the negatively charged ions of the protium ${}^{1}$H$^{-}$, deuterium 
${}^{2}$H$^{-}$ and tritium ${}^{3}$H$^{-}$ with these nuclear masses can be found in Table I of \cite{Fro2014}. Numerical computations of the photodetachment cross-sections of the negatively charged hydrogen
ions are more complicated than for the model ${}^{\infty}$H$^{-}$ ion. General theory of atomic photodetachment leads to a conclusion that photodetachment cross-section of a less bounded electron is always smaller 
than analogous cross-sections of strongly bounded atomic electrons. On the other hand, the negatively charged hydrogen ions have the $1s^2-$electron configuration. The 
less bounded electrons in this case mean larger `geometric radius' of the negatively charged hydrogen ion, i.e. a larger cross-section of photodetachment. In reality, there is a competition of these two opposite 
factors in the following series of ions: ${}^{\infty}$H$^{-}$, ${}^{3}$H$^{-}$ (tritium), ${}^{2}$H$^{-}$ (deuterium) and ${}^{1}$H$^{-}$ (protium).  

Let us discuss possible corrections to the photodetachment cross-section which are directly related to the finite nuclear mass. There are two groups of such corrections. The first group contains the common 
factor $1 + \frac{1}{M}$ (in atomic units), where $M$ is the mass of the central nucleus. These corrections arise due to additional factor(s) in the atomic wave functions of the final hydrogen atoms. The second 
group of corrections includes the common factor $1 + \frac{1}{M + 1}$. These corrections are directly related with the hydrogen atom recoil during photodetachment of the H$^{-}$ ion. For the negatively charged 
hydrogen ions the energy conservation law takes the form: $\hbar \omega = \frac{p^2_e}{2 m_e} + \frac{P^2_H}{2 (M + 1)} + I$, where $P_{H} = \mid {\bf P}_H \mid$ is the absolute value of momentum of the neutral 
hydrogen atom and $I$ is the ionization potential of the H$^{-}$ ion. The conservation of the electron and nuclear momenta in the H$^{-}$ ion leads to the relation: ${\bf p}_e = - {\bf P}_{\rm H}$, or $p_e = 
P_{\rm H}$. From these two equations one can write:
\begin{eqnarray}
     \hbar \omega = \frac{p^2_e}{2 m_e} \Bigl[ 1 + \frac{1}{2 (M + 1)} \Bigr] + I = \frac{p^2_e}{2} \Bigl[ 1 + \frac{1}{2 (M + 1)} \Bigr] + \frac{\gamma^2}{2} \label{energy}
\end{eqnarray}
where $\gamma = \sqrt{2 I}$ and equality in the right-hand side of Eq.(\ref{energy}) is written in atomic units. This lead to the following expression for the differential cross-section of the photodetachment 
of the H$^{-}$ ion \cite{Fro2014}:
\begin{eqnarray}
 & & d\sigma_{\nu} = \frac{1}{\pi} \alpha a^2_0 \Bigl[\frac{p_e}{p^2_e \bigl[1 + \frac{1}{2 (M + 1)}\bigr] + \gamma^2}\Bigr] \cdot \Bigl( 1 + \frac{1}{M+1} \Bigr)^2 \times \; \; \; \; \label{knu2} \\
 & & \Bigl| \Bigl\{\int \int \phi_{{\rm H}}(r_{31}) j_{1}(p_e r_{32}) \Bigl[ {\bf e}_f \cdot \Bigl(\frac{\partial}{\partial {\bf r}_1} + \frac{\partial}{\partial 
 {\bf r}_2}\Bigr) \Bigr] \Psi_{{\rm H^{-}}}(r_{32}, r_{31}, r_{21}) r_{32} r_{31} r_{21} dr_{32} dr_{31} dr_{21} d\Omega \Bigr\} \Bigr|^2 do \nonumber 
\end{eqnarray}
where the factor $\Bigl(1 + \frac{1}{M+1}\Bigr)^2$ reflects the fact that the electron velocity in the hydrogen atom/ion with the finite nuclear mass is larger than in analogous hydrogen atom with 
infinitely heavy nucleus. There is also an additional mass dependent factor $\Bigl(1 + \frac{1}{M}\Bigr)$ which is included in the normalization constant of the bound state wave function of the final hydrogen 
atom $\phi_{{\rm H}}(r_{31})$. In general, combination of a number of mass-dependent factors leads to relatively complex mass-dependence of the differential cross-section of the photodetachment of the H$^{-}$ 
ion. Results o four preliminary calculations of the photodetachment cross-sections of the ${}^{1}$H$^{-}$ (protium), ${}^{2}$H$^{-}$ (or D$^{-}$) and ${}^{3}$H$^{-}$ (or T$^{-}$) can be found in Table IV for 
different energies of the photo-electron $E_e = \frac{p^2_e}{2}$ (in atomic units). In calculations of each ion shown in Table IV we have used 350 basis functions in Eq.(\ref{exp}). Photodetachment cross-section
of the model ion ${}^{\infty}$H$^{-}$ (determined with 350 basis functions) is also presented in Table IV. The corresponding total energies obtained with these trial wave functions are also listed in Table IV.
Note that these energies are very close to the `exact' vales obtained in \cite{Fro2014}. This indicates very high quality of the trial wave functions. As follows from Table IV these cross-sections are relatively 
close to the photodetachment cross-sections of the H$^{-}$ ion (see Table II). However, some differences in such cross-sections do exist and must be investigated. We are planning to investigate such differences 
in our next study. 

\section{Conclusions}

We have considered the photodetachment of the negatively charged hydrogen ion(s). By using our new method we determine the photodetachment cross-sections of the H$^{-}$ ion with the infinitely heavy nucleus and 
for analogous ions of some hydrogen isotopes, including protium ${}^{1}$H$^{-}$, deuterium ${}^{2}$H$^{-}$ (or D$^{-}$) and tritium ${}^{3}$H$^{-}$ (or T$^{-}$). The crucial part of such calculations is highly
accurate computations of the three-particle integrals which also contain spherical Bessel functions of the first kind and truly correlated wave functions of the two-electron hydrogen ions H$^{-}$. In our method 
to solve all problems arising in this part we apply the Rayleigh's formula for the spherical Bessel function $j_{1}(p_e r_{32})$. Photodetachment cross-sections of the negatively charged ${}^{\infty}$H$^{-}$ 
ions determined in this study perfectly coincide with the photodetachment cross-sections determined in experiments \cite{PR1955A} - \cite{PRA2015}. Furthermore, the agreement between data from Table I in 
\cite{PRA2015} and our cross-sections determined at the same energies (see Tables I and II) is close to absolute (or 1:1). In general, our photodetachment cross-sections computed in the `velocity representation' 
are located between five experimental curves ploted in Fig.5 from \cite{PRA2015}. In fact, our data are very close to the experimental curve ploted by small circles `$\circ$', which correspond experimental data 
measured in \cite{PRA2015}. It should be mentioned that the paper \cite{PRA2015} was not known to me when I conducted these calculations (it was published after this paper has been submitted). Such an accurate 
coincidence indicates clearly the `velocity picture' used by us and based on the fundamental QED rules is appropriate and accurate to consider photodetachment of the negatively charged hydrogen ion(s). Moreover, 
a very accurate coincidence of our (theoretical) and experimental data shows that overall contribution of the electron-electron correlations and other similar effects into the H$^{-}$ photodetachemnt 
cross-sections is relatively small andcan be neglected in the first approximation (as expected) (see discussions in \cite{Fro2014} and \cite{PRA1994}). In part, such an accurate agreement with the experimental 
data can be expained by the fact that in our calculations we have applied very accurate bound state wave functions of the H$^{-}$ ion and its isotopes. For instance, the total energies of the ${}^{\infty}$H$^{-}$ 
ion obtained with 300 and 350 exponential basis functions are -0.527751016544180 $a.u.$ and -0.527751016544278 $a.u.$ which coincide with the `exact' energy (see, e.g., \cite{Fro2014}) in 13 decimal digits. 

In the case when the final hydrogen atom is formed in the excited $2^2S-$state we still do not have reliable experimental data, since any direct experimental study of the H$^{-}$ ion photodetachment with the 
formation of the final hydrogen atom in the excited $S-$states is extremely difficult. A few theoretical evaluations of the corresponding cross-section(s) \cite{PRA1994} - \cite{excite} were only approximate.  
In this study we consider photodetachment of the negatively charged hydrogen ions in those cases when the final hydrogen atom is formed in the excited $2s-$state. The corresponding cross-sections are determined 
with the use of numerical differentiation of the three-particle integral(s) in respect with one of its four parameters. The computed photodetachment cross-sections are in agreement with the cross-sections 
evaluated earlier in \cite{BD1} and \cite{excite}. However, we have to note again that the procedure of numerical differention was found to be sufficintly accurate for our current purposes, but in the future 
highly accurate calculations it must be replaced by more accurate methods. Briefly, we can say that our current results from Table III can be considered as preliminary. 

Another problem discussed in this study is to determine photodetachment cross-sections of the negatively charged, two-electron ions of the hydrogen isotopes: ${}^{1}$H$^{-}$ (protium), ${}^{2}$H$^{-}$ 
(deuterium D$^{-}$) and ${}^{3}$H$^{-}$ (tritium T$^{-}$). In each of these ions the nuclear masses are finite. Recently, we have made a significant progress in study of the photodetachment of such `real' 
hydrogen ions, but still a few additional steps are needed to obtain an absolute agreement with the known experimental data. In particular, we need to take into account all effect related with 
electron-electron correlations in the final state of the photodetachment process, Eq.(\ref{e0}). In general, such corrections are small, but their overall contribution can noticeably change cross-sections at 
some energies. Our approach to include such corrections is described in detail in \cite{Fro2014}. Analysis of such corrections will be considered in the next study. Another approach to this problem was 
discussed in \cite{Bhat2}.

\newpage
\begin{table}[tbp]
   \caption{Photodetachment cross-section (in $cm^{2}$) of the negatively charged ${}^{\infty}$H$^{-}$ ion in the ground $1^1S-$state. 
           $N$ is the total number of basis function in the trial wave function and $E$ is the total energy (in $a.u.$) for this wave 
           function. The final hydrogen atom is formed in the ground $1s$-state. The notation $E_e$ stands for the energy of the emitted 
           photo-electron (in $a.u.$).}
     \begin{center}
     \scalebox{0.70}{%
     \begin{tabular}{| c | c | c | c | c | c |}
      \hline\hline
  $E_{{\rm H}^{-}}$  & -0.5277751015678648 & -0.5277751016514754 & -0.527751016542222 & -0.527751016544180 & -0.527751016544278 \\ 
           \hline
  $E_e$  & $\sigma(E_e) (N$ = 50) & $\sigma(E_e) (N$ = 100) & $\sigma(E_e) (N$ = 200) & $\sigma(E_e) (N$ = 300) & $\sigma(E_e) (N$ = 350) \\
          \hline    
   0.0000125  & 0.29605783690E-17 &  0.29598715990E-17 &  0.29596998756E-17 &  0.29598550732E-17 &  0.29598566799E-17 \\
   0.0001125  & 0.87973761028E-17 &  0.87953808289E-17 &  0.87949302746E-17 &  0.87953448276E-17 &  0.87953664283E-17 \\
   0.0003125  & 0.14386547859E-16 &  0.14383603423E-16 &  0.14382993623E-16 &  0.14383551527E-16 &  0.14383597731E-16 \\
   0.0006125  & 0.19580556054E-16 &  0.19577122883E-16 &  0.19576430332E-16 &  0.19577016434E-16 &  0.19577077184E-16 \\
   0.0010125  & 0.24256110179E-16 &  0.24252654008E-16 &  0.24251915406E-16 &  0.24252450461E-16 &  0.24252520229E-16 \\
   0.0015125  & 0.28320090502E-16 &  0.28316982705E-16 &  0.28316242717E-16 &  0.28316688061E-16 &  0.28316758397E-16 \\
   0.0021125  & 0.31712467227E-16 &  0.31709927491E-16 &  0.31709256092E-16 &  0.31709607249E-16 &  0.31709665345E-16 \\
   0.0028125  & 0.34406250159E-16 &  0.34404338215E-16 &  0.34403814787E-16 &  0.34404089266E-16 &  0.34404125710E-16 \\
   0.0036125  & 0.36404882768E-16 &  0.36403529204E-16 &  0.36403217986E-16 &  0.36403439874E-16 &  0.36403454405E-16 \\
   0.0045125  & 0.37737746241E-16 &  0.37736804077E-16 &  0.37736736862E-16 &  0.37736924254E-16 &  0.37736925369E-16 \\
                \hline
   0.0055125  & 0.38454542597E-16 &  0.38453841338E-16 &  0.38454011788E-16 &  0.38454171384E-16 &  0.38454171955E-16 \\
   0.0066125  & 0.38619295849E-16 &  0.38618684920E-16 &  0.38619053036E-16 &  0.38619181632E-16 &  0.38619193960E-16 \\
   0.0078125  & 0.38304584444E-16 &  0.38303959740E-16 &  0.38304462825E-16 &  0.38304552518E-16 &  0.38304585002E-16 \\
   0.0091125  & 0.37586442834E-16 &  0.37585756010E-16 &  0.37586322075E-16 &  0.37586365790E-16 &  0.37586421743E-16 \\
   0.0105125  & 0.36540186616E-16 &  0.36539440563E-16 &  0.36540000770E-16 &  0.36539996014E-16 &  0.36540074168E-16 \\
   0.0120125  & 0.35237254090E-16 &  0.35236489739E-16 &  0.35236987673E-16 &  0.35236937881E-16 &  0.35237033798E-16 \\
   0.0136125  & 0.33743033576E-16 &  0.33742313147E-16 &  0.33742710013E-16 &  0.33742623905E-16 &  0.33742731565E-16 \\
   0.0153125  & 0.32115564584E-16 &  0.32114955391E-16 &  0.32115231333E-16 &  0.32115121065E-16 &  0.32115234221E-16 \\
   0.0171125  & 0.30404958234E-16 &  0.30404519702E-16 &  0.30404672073E-16 &  0.30404551036E-16 &  0.30404664124E-16 \\
   0.0190125  & 0.28653369842E-16 &  0.28653144950E-16 &  0.28653184692E-16 &  0.28653065525E-16 &  0.28653174140E-16 \\
             \hline
   0.0210125  & 0.26895364963E-16 &  0.26895376194E-16 &  0.26895323402E-16 &  0.26895216560E-16 &  0.26895317595E-16 \\
   0.0231125  & 0.25158540750E-16 &  0.25158789884E-16 &  0.25158669419E-16 &  0.25158582386E-16 &  0.25158673936E-16 \\
   0.0253125  & 0.23464290471E-16 &  0.23464761154E-16 &  0.23464598890E-16 &  0.23464535980E-16 &  0.23464617137E-16 \\
   0.0276125  & 0.21828625603E-16 &  0.21829287584E-16 &  0.21829107569E-16 &  0.21829070170E-16 &  0.21829140775E-16 \\
   0.0300125  & 0.20262994215E-16 &  0.20263808223E-16 &  0.20263630834E-16 &  0.20263617929E-16 &  0.20263678346E-16 \\
   0.0325125  & 0.18775054831E-16 &  0.18775977338E-16 &  0.18775818331E-16 &  0.18775827151E-16 &  0.18775878074E-16 \\
   0.0351125  & 0.17369381282E-16 &  0.17370368625E-16 &  0.17370238921E-16 &  0.17370265606E-16 &  0.17370307915E-16 \\
   0.0378125  & 0.16048086277E-16 &  0.16049097863E-16 &  0.16049003856E-16 &  0.16049044055E-16 &  0.16049078706E-16 \\
   0.0406125  & 0.14811360133E-16 &  0.14812360646E-16 &  0.14812304846E-16 &  0.14812354194E-16 &  0.14812382158E-16 \\
   0.0435125  & 0.13657926934E-16 &  0.13658887600E-16 &  0.13658869441E-16 &  0.13658923911E-16 &  0.13658946126E-16 \\
            \hline
   0.0465125  & 0.12585423999E-16 &  0.12586323045E-16 &  0.12586339722E-16 &  0.12586395851E-16 &  0.12586413199E-16 \\
   0.0496125  & 0.11590712460E-16 &  0.11591534944E-16 &  0.11591582203E-16 &  0.11591637205E-16 &  0.11591650497E-16 \\
   0.0528125  & 0.10670127572E-16 &  0.10670864770E-16 &  0.10670937586E-16 &  0.10670989381E-16 &  0.10670999349E-16 \\
   0.0561125  & 0.98196773647E-17 &  0.98203258858E-17 &  0.98204190031E-17 &  0.98204661807E-17 &  0.98204734767E-17 \\
   0.0595125  & 0.90351977918E-17 &  0.90357585655E-17 &  0.90358669036E-17 &  0.90359086434E-17 &  0.90359138388E-17 \\
   0.0630125  & 0.83124717978E-17 &  0.83129490365E-17 &  0.83130679640E-17 &  0.83131039366E-17 &  0.83131075229E-17 \\
   0.0666125  & 0.76473188426E-17 &  0.76477190671E-17 &  0.76478445644E-17 &  0.76478748231E-17 &  0.76478772150E-17 \\
   0.0703125  & 0.70356605505E-17 &  0.70359917250E-17 &  0.70361204564E-17 &  0.70361453329E-17 &  0.70361468723E-17 \\
   0.0741125  & 0.64735672865E-17 &  0.64738380909E-17 &  0.64739674087E-17 &  0.64739874186E-17 &  0.64739883797E-17 \\ 
   0.0780125  & 0.59572896770E-17 &  0.59575089258E-17 &  0.59576368268E-17 &  0.59576525896E-17 &  0.59576531852E-17 \\ 
    \hline\hline
  \end{tabular}}
  \end{center}
  \end{table}
\newpage
\begin{table}[tbp]
   \caption{Photodetachment cross-section (in $cm^{2}$) of the negatively charged hydrogen ion ${}^{\infty}$H$^{-}$ 
            in its ground $1^1S-$states. The final hydrogen atom is formed in the ground $s-$state.}
     \begin{center}
     \scalebox{0.70}{%
     \begin{tabular}{| c | c | c | c | c | c | c | c | c | c |}
      \hline\hline
 $i$ & $\sigma(E_e)$ & $i$ & $\sigma(E_e)$ & $i$ & $\sigma(E_e)$ & $i$ & $\sigma(E_e)$ & $i$ & $\sigma(E_e)$ \\
          \hline    
   1 & 0.29598566799E-17 & 41 & 0.26895317595E-16 &  81 & 0.54835922121E-17 & 121 & 0.11299827712E-17 & 161 & 0.27970318521E-18 \\
   2 & 0.58985789927E-17 & 42 & 0.26022872407E-16 &  82 & 0.52613628442E-17 & 122 & 0.10887526835E-17 & 162 & 0.27072608218E-18 \\
   3 & 0.87953664283E-17 & 43 & 0.25158673936E-16 &  83 & 0.50484667466E-17 & 123 & 0.10491519452E-17 & 163 & 0.26206475142E-18 \\
   4 & 0.11630086856E-16 & 44 & 0.24305206233E-16 &  84 & 0.48445201077E-17 & 124 & 0.10111119507E-17 & 164 & 0.25370717434E-18 \\
   5 & 0.14383597731E-16 & 45 & 0.23464617137E-16 &  85 & 0.46491524785E-17 & 125 & 0.97456711729E-18 & 165 & 0.24564182171E-18 \\
   6 & 0.17038037341E-16 & 46 & 0.22638744308E-16 &  86 & 0.44620066437E-17 & 126 & 0.93945474920E-18 & 166 & 0.23785763256E-18 \\
   7 & 0.19577077184E-16 & 47 & 0.21829140775E-16 &  87 & 0.42827384445E-17 & 127 & 0.90571490759E-18 & 167 & 0.23034399406E-18 \\
   8 & 0.21986131388E-16 & 48 & 0.21037099731E-16 &  88 & 0.41110165611E-17 & 128 & 0.87329028622E-18 & 168 & 0.22309072227E-18 \\
   9 & 0.24252520229E-16 & 49 & 0.20263678346E-16 &  89 & 0.39465222618E-17 & 129 & 0.84212609272E-18 & 169 & 0.21608804378E-18 \\
  10 & 0.26365585671E-16 & 50 & 0.19509720468E-16 &  90 & 0.37889491257E-17 & 130 & 0.81216993515E-18 & 170 & 0.20932657818E-18 \\
         \hline
  11 & 0.28316758397E-16 & 51 & 0.18775878074E-16 &  91 & 0.36380027431E-17 & 131 & 0.78337171363E-18 & 171 & 0.20279732129E-18 \\
  12 & 0.30099577464E-16 & 52 & 0.18062631434E-16 &  92 & 0.34934003992E-17 & 132 & 0.75568351683E-18 & 172 & 0.19649162916E-18 \\
  13 & 0.31709665345E-16 & 53 & 0.17370307915E-16 &  93 & 0.33548707456E-17 & 133 & 0.72905952322E-18 & 173 & 0.19040120278E-18 \\
  14 & 0.33144662384E-16 & 54 & 0.16699099457E-16 &  94 & 0.32221534618E-17 & 134 & 0.70345590671E-18 & 174 & 0.18451807344E-18 \\
  15 & 0.34404125710E-16 & 55 & 0.16049078706E-16 &  95 & 0.30949989111E-17 & 135 & 0.67883074661E-18 & 175 & 0.17883458886E-18 \\
  16 & 0.35489398315E-16 & 56 & 0.15420213842E-16 &  96 & 0.29731677933E-17 & 136 & 0.65514394168E-18 & 176 & 0.17334339972E-18 \\
  17 & 0.36403454405E-16 & 57 & 0.14812382158E-16 &  97 & 0.28564307959E-17 & 137 & 0.63235712812E-18 & 177 & 0.16803744706E-18 \\
  18 & 0.37150727217E-16 & 58 & 0.14225382425E-16 &  98 & 0.27445682467E-17 & 138 & 0.61043360131E-18 & 178 & 0.16290994995E-18 \\
  19 & 0.37736925369E-16 & 59 & 0.13658946126E-16 &  99 & 0.26373697687E-17 & 139 & 0.58933824107E-18 & 179 & 0.15795439398E-18 \\
  20 & 0.38168843436E-16 & 60 & 0.13112747602E-16 & 100 & 0.25346339389E-17 & 140 & 0.56903744043E-18 & 180 & 0.15316452004E-18 \\
          \hline
  21 & 0.38454171955E-16 & 61 & 0.12586413199E-16 & 101 & 0.24361679531E-17 & 141 & 0.54949903762E-18 & 181 & 0.14853431369E-18 \\
  22 & 0.38601311401E-16 & 62 & 0.12079529466E-16 & 102 & 0.23417872957E-17 & 142 & 0.53069225118E-18 & 182 & 0.14405799505E-18 \\
  23 & 0.38619193960E-16 & 63 & 0.11591650497E-16 & 103 & 0.22513154172E-17 & 143 & 0.51258761806E-18 & 183 & 0.13973000897E-18 \\
  24 & 0.38517116206E-16 & 64 & 0.11122304450E-16 & 104 & 0.21645834193E-17 & 144 & 0.49515693459E-18 & 184 & 0.13554501582E-18 \\
  25 & 0.38304585002E-16 & 65 & 0.10670999349E-16 & 105 & 0.20814297481E-17 & 145 & 0.47837320012E-18 & 185 & 0.13149788255E-18 \\
  26 & 0.37991178291E-16 & 66 & 0.10237228196E-16 & 106 & 0.20016998954E-17 & 146 & 0.46221056329E-18 & 186 & 0.12758367418E-18 \\
  27 & 0.37586421743E-16 & 67 & 0.98204734767E-17 & 107 & 0.19252461090E-17 & 147 & 0.44664427071E-18 & 187 & 0.12379764566E-18 \\
  28 & 0.37099681674E-16 & 68 & 0.94202111004E-17 & 108 & 0.18519271127E-17 & 148 & 0.43165061812E-18 & 188 & 0.12013523411E-18 \\
  29 & 0.36540074168E-16 & 69 & 0.90359138388E-17 & 109 & 0.17816078341E-17 & 149 & 0.41720690363E-18 & 189 & 0.11659205135E-18 \\
  30 & 0.35916389888E-16 & 70 & 0.86670543053E-17 & 110 & 0.17141591424E-17 & 150 & 0.40329138327E-18 & 190 & 0.11316387675E-18 \\
          \hline
  31 & 0.35237033798E-16 & 71 & 0.83131075229E-17 & 111 & 0.16494575956E-17 & 151 & 0.38988322848E-18 & 191 & 0.10984665047E-18 \\
  32 & 0.34509978723E-16 & 72 & 0.79735531232E-17 & 112 & 0.15873851957E-17 & 152 & 0.37696248558E-18 & 192 & 0.10663646689E-18 \\
  33 & 0.33742731565E-16 & 73 & 0.76478772150E-17 & 113 & 0.15278291539E-17 & 153 & 0.36451003711E-18 & 193 & 0.10352956835E-18 \\
  34 & 0.32942310842E-16 & 74 & 0.73355739575E-17 & 114 & 0.14706816644E-17 & 154 & 0.35250756493E-18 & 194 & 0.10052233922E-18 \\
  35 & 0.32115234221E-16 & 75 & 0.70361468723E-17 & 115 & 0.14158396867E-17 & 155 & 0.34093751501E-18 & 195 & 0.97611300126E-19 \\
  36 & 0.31267514701E-16 & 76 & 0.67491099212E-17 & 116 & 0.13632047363E-17 & 156 & 0.32978306387E-18 & 196 & 0.94793102516E-19 \\
  37 & 0.30404664124E-16 & 77 & 0.64739883797E-17 & 117 & 0.13126826846E-17 & 157 & 0.31902808646E-18 & 197 & 0.92064523376E-19 \\
  38 & 0.29531702809E-16 & 78 & 0.62103195285E-17 & 118 & 0.12641835659E-17 & 158 & 0.30865712563E-18 & 198 & 0.89422460223E-19 \\
  39 & 0.28653174140E-16 & 79 & 0.59576531852E-17 & 119 & 0.12176213929E-17 & 159 & 0.29865536289E-18 & 199 & 0.86863926295E-19 \\
  40 & 0.27773163068E-16 & 80 & 0.57155520971E-17 & 120 & 0.11729139794E-17 & 160 & 0.28900859057E-18 & 200 & 0.84386045937E-19 \\
    \hline\hline
  \end{tabular}}
  \end{center}
  \end{table}
\newpage
\begin{table}[tbp]
   \caption{Photodetachment cross-section (in $cm^{2}$) of the negatively charged hydrogen ion ${}^{\infty}$H$^{-}$ 
            in its ground $1^1S-$states. The final hydrogen atom is formed in the first excited $s-$state.}
     \begin{center}
     \scalebox{0.70}{%
     \begin{tabular}{| c | c | c | c | c | c | c | c |}
      \hline\hline
 $p_e$ & $\sigma(E_e)$ & $p_e$ & $\sigma(E_e)$ & $p_e$ & $\sigma(E_e)$ & $p_e$ & $\sigma(E_e)$ \\
          \hline    
  0.01 &  0.15169453769E-18 & 0.16 &  0.16689095636E-17 & 0.31 &  0.21171774945E-17 & 0.46 &  0.18258475296E-17 \\
  0.02 &  0.29513081491E-18 & 0.17 &  0.17180044561E-17 & 0.32 &  0.21231650513E-17 & 0.47 &  0.17848013982E-17 \\
  0.03 &  0.42950207831E-18 & 0.18 &  0.17632152811E-17 & 0.33 &  0.21252205995E-17 & 0.48 &  0.17424897768E-17 \\
  0.04 &  0.55759395646E-18 & 0.19 &  0.18052224207E-17 & 0.34 &  0.21232787756E-17 & 0.49 &  0.16991672981E-17 \\
  0.05 &  0.68210106756E-18 & 0.20 &  0.18445390636E-17 & 0.35 &  0.21173257734E-17 & 0.50 &  0.16550750965E-17 \\
  0.06 &  0.80402375798E-18 & 0.21 &  0.18815119924E-17 & 0.36 &  0.21073985814E-17 & 0.51 &  0.16104391465E-17 \\
  0.07 &  0.92268403351E-18 & 0.22 &  0.19163339790E-17 & 0.37 &  0.20935824448E-17 & 0.52 &  0.15654691307E-17 \\
  0.08 &  0.10364910897E-17 & 0.23 &  0.19490634468E-17 & 0.38 &  0.20760069165E-17 & 0.53 &  0.15203577762E-17 \\
  0.09 &  0.11437296099E-17 & 0.24 &  0.19796480642E-17 & 0.39 &  0.20548408558E-17 & 0.54 &  0.14752805973E-17 \\
  0.10 &  0.12430738115E-17 & 0.25 &  0.20079498256E-17 & 0.40 &  0.20302867197E-17 & 0.55 &  0.14303959796E-17 \\
  0.11 &  0.13338055308E-17 & 0.26 &  0.20337698926E-17 & 0.41 &  0.20025744610E-17 & 0.56 &  0.13858455425E-17 \\
  0.12 &  0.14158199181E-17 & 0.27 &  0.20568720138E-17 & 0.42 &  0.19719553114E-17 & 0.57 &  0.13417547244E-17 \\
  0.13 &  0.14895139123E-17 & 0.28 &  0.20770037429E-17 & 0.43 &  0.19386956840E-17 & 0.58 &  0.12982335353E-17 \\
  0.14 &  0.15556298503E-17 & 0.29 &  0.20939149690E-17 & 0.44 &  0.19030713856E-17 & 0.59 &  0.12553774311E-17 \\
  0.15 &  0.16150991726E-17 & 0.30 &  0.21073734863E-17 & 0.45 &  0.18653622838E-17 & 0.60 &  0.12132682685E-17 \\
      \hline\hline
  \end{tabular}}
  \end{center}
  \end{table}
\newpage
\begin{table}[tbp]
   \caption{Photodetachment cross-section (in $cm^{2}$) of the negatively charged hydrogen ion in their ground $1^1S-$states. 
           $N$ is the total number of basis function in the trial wave function and $E$ is the total energy (in $a.u.$) for this wave 
           function. The final hydrogen atom is formed in the ground $1s$-state. The notation $E_e$ stands for the kinetic energy of 
           the emitted photo-electron (in $a.u.$).}
     \begin{center}
     \scalebox{0.75}{%
     \begin{tabular}{| c | c | c | c | c |}
      \hline\hline
        & ${}^{\infty}$H$^{-}$ & ${}^{3}$H$^{-}$ & ${}^{2}$H$^{-}$ & ${}^{1}$H$^{-}$ \\
           \hline
  $E_{{\rm H}^{-}}$  & -0.5277751016544278 & -0.527649048201840 & -0.527598324689626 & -0.527445881119668 \\ 
           \hline
  $E_e$  & $\sigma(E_e) (N$ = 350) & $\sigma(E_e) (N$ = 350) & $\sigma(E_e) (N$ = 350) & $\sigma(E_e) (N$ = 350) \\
          \hline     
 0.0000500  & 0.2959856679870E-17 &  0.5898071851203E-17 &  0.5897932817961E-17 & 0.58970881335E-17 \\
 0.0001125  & 0.5898578992686E-17 &  0.8794584542552E-17 &  0.8794335668102E-17 & 0.87930359069E-17 \\
 0.0002000  & 0.8795366428292E-17 &  0.1162901145802E-16 &  0.1162861725426E-16 & 0.11626832132E-16 \\
 0.0003125  & 0.1163008685588E-16 &  0.1438220490456E-16 &  0.1438163186360E-16 & 0.14379323906E-16 \\
 0.0004500  & 0.1438359773146E-16 &  0.1703629503788E-16 &  0.1703551495594E-16 & 0.17032638389E-16 \\
 0.0006125  & 0.1703803734141E-16 &  0.1957494491642E-16 &  0.1957393521732E-16 & 0.19570436466E-16 \\
 0.0008000  & 0.1957707718383E-16 &  0.2198356259143E-16 &  0.2198230472500E-16 & 0.21978124069E-16 \\
 0.0010125  & 0.2198613138832E-16 &  0.2424946573180E-16 &  0.2424794299533E-16 & 0.24243016855E-16 \\
 0.0012500  & 0.2425252022884E-16 &  0.2636199692909E-16 &  0.2636019272279E-16 & 0.26354456186E-16 \\
 0.0015125  & 0.2636558567114E-16 &  0.2831258986430E-16 &  0.2831048683222E-16 & 0.28303876164E-16 \\
               \hline
 0.0018000  & 0.2831675839673E-16 &  0.3009478807359E-16 &  0.3009236812510E-16 & 0.30084823091E-16 \\
 0.0021125  & 0.3009957746377E-16 &  0.3170421932053E-16 &  0.3170146428578E-16 & 0.31692930205E-16 \\
 0.0024500  & 0.3170966534484E-16 &  0.3313852966036E-16 &  0.3313542227316E-16 & 0.33125851743E-16 \\
 0.0028125  & 0.3314466238433E-16 &  0.3439728216636E-16 &  0.3439380711190E-16 & 0.34383161403E-16 \\
 0.0032000  & 0.3440412571010E-16 &  0.3548182594435E-16 &  0.3547797079138E-16 & 0.35466220892E-16 \\
 0.0036125  & 0.3548939831482E-16 &  0.3639514145850E-16 &  0.3639089737677E-16 & 0.36378024707E-16 \\
 0.0040500  & 0.3640345440464E-16 &  0.3714166832346E-16 &  0.3713703055060E-16 & 0.37123027354E-16 \\
 0.0045125  & 0.3715072721740E-16 &  0.3772712159963E-16 &  0.3772208966385E-16 & 0.37706959091E-16 \\
 0.0050000  & 0.3773692536921E-16 &  0.3815830229366E-16 &  0.3815288002287E-16 & 0.38136635946E-16 \\
 0.0055125  & 0.3816884343565E-16 &  0.3844290725784E-16 &  0.3843710261617E-16 & 0.38419769197E-16 \\
               \hline 
 0.0060500  & 0.3845417195483E-16 &  0.3858934304742E-16 &  0.3858316784240E-16 & 0.38564778897E-16 \\
 0.0066125  & 0.3860131140057E-16 &  0.3860654758202E-16 &  0.3860001708227E-16 & 0.38580615273E-16 \\
 0.0072000  & 0.3861919396038E-16 &  0.3850382270889E-16 &  0.3849695520654E-16 & 0.38476591109E-16 \\
 0.0078125  & 0.3851711620594E-16 &  0.3829068002096E-16 &  0.3828349636687E-16 & 0.38262227446E-16 \\
 0.0084500  & 0.3830458500200E-16 &  0.3797670157192E-16 &  0.3796922470517E-16 & 0.37947114237E-16 \\
 0.0091125  & 0.3799117829127E-16 &  0.3757141647769E-16 &  0.3756367096565E-16 & 0.37540786939E-16 \\
 0.0098000  & 0.3758642174276E-16 &  0.3708419381512E-16 &  0.3707620541543E-16 & 0.37052619444E-16 \\
 0.0105125  & 0.3709968167445E-16 &  0.3652415173423E-16 &  0.3651594698662E-16 & 0.36491733251E-16 \\
 0.0112500  & 0.3654007416812E-16 &  0.3590008229333E-16 &  0.3589168814598E-16 & 0.35866922392E-16 \\
 0.0120125  & 0.3591638988830E-16 &  0.3522039120618E-16 &  0.3521183467941E-16 & 0.35186593282E-16 \\
       \hline\hline
  \end{tabular}}
  \end{center}
  \end{table}
\end{document}